\documentclass[journal=jpclcd,manuscript=letter,layout=twocolumn]{achemso}
\usepackage{graphicx}
\usepackage{amsmath,amssymb}
\usepackage{dcolumn}
\usepackage{bm}
\usepackage{extarrows}
\usepackage{xurl}
\usepackage{makecell}
\usepackage{bbding}

\makeatletter
\renewcommand*{\acs@abstract@extras}{%
  \acs@keywords@print
  \newpage
}
\makeatother

\title{A Mesoscopic Ginzburg--Landau Model for Vibrational Strong Coupling Enhanced Rayleigh Scattering in Molecular Liquids}

\author{Wenxiang Ying}
\email{wying3@sas.upenn.edu}
\affiliation{Department of Chemistry, University of Pennsylvania, Philadelphia, Pennsylvania 19104, USA}

\author{Abraham Nitzan}
\email{anitzan@sas.upenn.edu}
\affiliation{Department of Chemistry, University of Pennsylvania, Philadelphia, Pennsylvania 19104, USA}
\alsoaffiliation{School of Chemistry, Tel Aviv University, Tel Aviv 69978, Israel}

\begin{document}

\begin{abstract}
Recent experiments by Sandeep \textit{et al.} [Angew. Chem. Int. Ed. 65, e16917 (2026)] suggest that vibrational strong coupling (VSC) in molecular liquids can generate mesoscopic phenomena beyond single-molecule observables, including resonantly enhanced Rayleigh scattering, abrupt concentration thresholds, and thermal collapse. Motivated by these observations, we construct a mesoscopic Ginzburg--Landau model with two coupled fields: a cavity-controlled collective vibrational polarization $P$ and a secondary structural field $m$ whose long-wavelength susceptibility is renormalized by the collective vibrational polarization intensity $P^2$, assumed to govern long-wavelength density/dielectric fluctuations. With calibrated parameters, the model captures the observed Rayleigh enhancement, collective scaling relations, and threshold-like behavior, while explaining why polaritonic/IR signatures may persist when Rayleigh scattering disappears.  The model further predicts enhanced long-wavelength density/dielectric correlations, enlarged mesoscopic correlation lengths, and slowed structural dynamics in the regime with strong Rayleigh enhancement, providing direct experimental tests through small-angle X-ray/neutron scattering and dynamic light-scattering probes.
\end{abstract}

\begin{figure}[H]
 \centering
 \begin{minipage}[t]{1.0\linewidth}
    \centering
    \includegraphics[width=\linewidth]{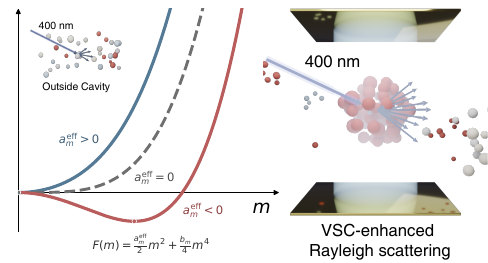}
    \caption*{TOC Graphic}
 \end{minipage}
\end{figure}

\newpage


Vibrational strong coupling (VSC) arises when an infrared-active molecular vibration
hybridizes with an optical cavity mode~\cite{Simpkins_Chemrev2023, Arkajit_Chemrev_2023, Hirai_ChemRev2023, Xiong_ChemRev2024}, producing vibrational polaritons
separated by the collective Rabi splitting $\Omega$.
Experiments have reported VSC-induced changes in ground-state reactivity, thermodynamic
parameters, stereoselectivity, enzymatic activity, and condensed-phase chemistry~\cite{Ebbesen_angew_2016,Ebbesen_angew_2019, Ebbesen_science_2019,Simpkins2023, Ebbesen_nanophotonics_2020, Ebbesen_Angew_2021, Lather_2022, Hirai_2020, verdelli_2024, Patrahau_Angew2024}.
These observations share several empirical features: resonance (maximum effect at normal incidence) with a molecular vibration,
collective scaling with molecular density or Rabi splitting, and modified chemical reaction rates in the absence of external optical pumping (thermal activation)~\cite{Wang2021AP,sidler2022JCP,Joel_VSC_Rev_2023, Ying_ARPC2026}. They have motivated extensive theoretical work on understanding polaritonic chemistry and
ground-state rate modification~\cite{Li_2021, JCao_2021,Arkajit_JCP_2022,Sun2022JPCL,Arkajit_jpcl2022,Schaefer2022,Wang2022JPCL, Wang2022Collective,Fischer2022JCP, Joel_prl_2022,Joel_PGH2023,Jeremy_RPMD_2023,Limmer_2023,Arkajit_2022,Ying2023,Hu2023,Ying2023NanoP,Ying2024,Ke_2024_1, Ke_2024_2,Reichman_Nanoph2024,Vega_Ying_Huo_2024,Fiechter_arXiv2026}.

\begin{figure}[htbp]
  \centering
  \includegraphics[width=\columnwidth]{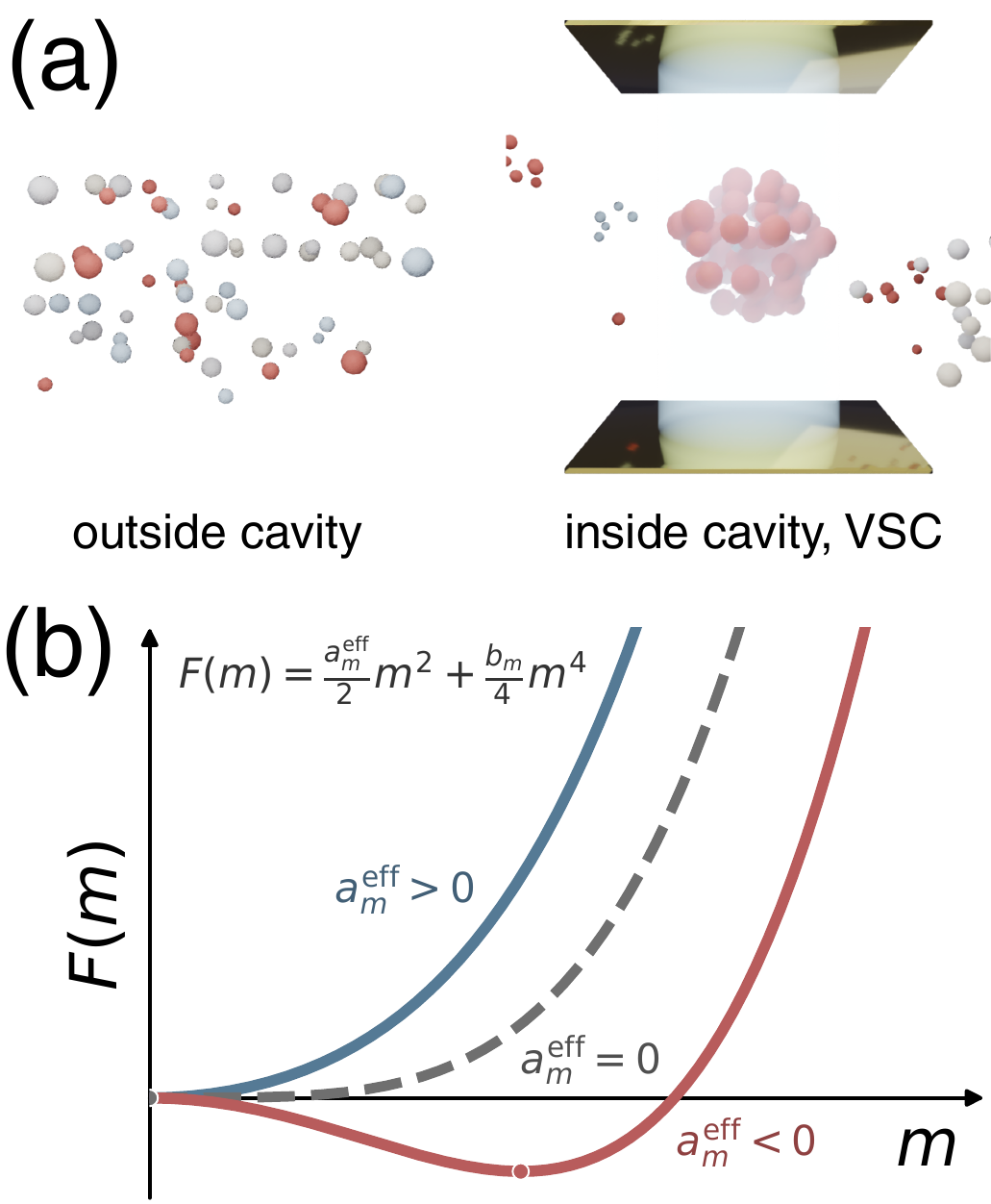}
  \caption{%
    \textbf{VSC-associated clustering as a mesoscopic Ginzburg--Landau instability.}
    (a) Schematic comparison of a dispersed molecular ensemble outside the cavity and a clustered mesoscopic state formed inside a Fabry--P\'erot cavity under VSC.
    (b) Schematic Landau free-energy landscape for the structural order parameter $m$, written in terms of an effective inverse susceptibility $a_m^{\rm eff}$ (Eq.~\ref{eq:aeff}) and $F(m)=a_m^{\rm eff}m^2/2+b_m m^4/4$. In the normal liquid, $a_m^{\rm eff}>0$ stabilizes the dispersed state at $m=0$. The $-\frac{g}{2}m^2P^2$ coupling reduces the structural stiffness from $a_m$ to $a_m^{\rm eff}=a_m-gP^2$ as $P^2$ grows inside the ordered-$P$ region ($a_P<0$), enhancing long-wavelength structural fluctuations. At the structural spinodal $a_m^{\rm eff}=0$ ($gP^2=a_m$), the $m=0$ branch loses its positive quadratic curvature; upon crossing into $a_m^{\rm eff}<0$, the minimum shifts continuously to nonzero $m$, signaling spontaneous structural ordering.
  }
  \label{fig:concept}
\end{figure}

The central unresolved issue is the collective effect of the modified chemical response~\cite{Joel_VSC_Rev_2023, Ying_ARPC2026}.
In the thermodynamic limit, where $N \sim 10^6 - 10^{12}$ molecules share a finite
collective Rabi splitting $\Omega$, the per-molecule coupling
$g_\mathrm{c} \propto \Omega/\sqrt{N}$ is small; a purely single-molecule
mechanism is therefore strongly diluted. 
Compounding this theoretical puzzle, previous claims of cavity-modified ground-state reactivity have sparked ongoing debate, as \textit{in-situ} infrared transmission measurements can be susceptible to macroscopic optical artifacts that convolute the kinetic analysis~\cite{Imperatore2021}. To isolate genuine collective polaritonic phenomena from such spectroscopic interference, rigorous measurements based on ex-situ chemical analysis or non-resonant optical measurements are strongly desired. 
Recent non-resonant light-scattering experiments by Sandeep \textit{et al.}~\cite{Sandeep2024} sharpen this issue. In their experiments, the 400~nm Rayleigh probe is non-resonant with the infrared cavity, while an infrared Fabry--Pérot cavity mode is tuned at normal incidence into resonance with a molecular vibrational band, such as the toluene C=C stretch at 1603~cm$^{-1}$, the toluene C--H stretch at 3027~cm$^{-1}$, or the broad water O--H stretch near 3400~cm$^{-1}$. Inside such on-resonance cavities, pure toluene and water show Rayleigh scattering enhanced by one to two orders of magnitude relative to off-resonance or uncoupled controls; the enhancement follows the vibrational resonance upon detuning, collapses abruptly upon dilution in toluene mixtures even while spectroscopic strong coupling persists, and disappears above a sharp transition temperature.
This threshold-like
onset and thermal collapse resemble a phase-transition-like change in a disordered
molecular liquid. Since small-wavevector Rayleigh scattering probes long-wavelength density/dielectric fluctuations rather than intramolecular dynamics, these observations indicate a VSC-associated mesoscopic structural response, suggesting a cluster formation picture. Fig.~\ref{fig:concept}a
schematizes this interpretation: outside the cavity the liquid remains a dispersed
ensemble, whereas resonant VSC is associated with mesoscopic clustered domains
that produce the enhanced Rayleigh signal.

Here we construct a mesoscopic Ginzburg--Landau (GL) model~\cite{GinzburgLandau1950, LandauLifshitz1980, Chaikin_Lubensky_1995} for this collective structural response, with a primary cavity-controlled collective vibrational polarization $P$ (see definition below), and a secondary mesoscopic structural field $m$ whose long-wavelength structural susceptibility is renormalized by the collective vibrational intensity $P^2$.
With calibrated parameters, the model captures Rayleigh scattering enhancement, collective scaling relations, and phase-transition-like behavior similar to that observed by Sandeep \textit{et al.}~\cite{Sandeep2024}, and suggests a two-step hierarchy linking polaritonic/IR signatures to a mesoscopic structural response.
It further predicts enhanced low-$q$ density/dielectric correlations and slowed mesoscopic structural dynamics in the structurally ordered regime, which can be tested by small-angle X-ray/neutron scattering, dynamic light scattering, and low-frequency dielectric or THz spectroscopy.

\section{Theoretical Model and Assumptions}
To begin with, we introduce the Ginzburg-Landau model, establish the mesoscopic coupling mechanism between the vibrational polarization and structural density fields under symmetry constraints, and introduce the phenomenological parametrization used to bridge the theoretical fields with real-space experimental signals. 

\subsection{Order parameters}
Guided by the instability sketched in Fig.~\ref{fig:concept}b which aims to account for the experimental observations, we propose a coarse-grained model characterized by two fields that capture the long-wavelength density/dielectric fluctuations at the mesoscopic level. These fields are as follows.

\paragraph{Collective vibrational polarization $P$.}
$P$ is the amplitude of the cavity-selected bright vibrational mode: the collective IR-active superposition
\mbox{$P \propto N^{-1/2}\sum_i \mu_i(x_i)$}
of molecular transition dipoles aligned with the cavity polarization axis. Here, we refer this collective vibrational response to the cavity field as ``polarization''.
$P$ is the \emph{primary} cavity-controlled variable, promoted by the collective Rabi splitting $\Omega$ and suppressed by detuning $\Delta = \omega_\mathrm{c} - \nu$ and by thermal fluctuations (measured by temperature $T$).
A finite $P$ therefore represents a collective vibrational response in the IR-active bright mode and in the cavity--vibration hybrid polariton modes derived from it. Its onset should be visible in IR and multidimensional IR spectroscopies through normal-mode splitting, oscillator-strength redistribution, linewidth changes, and spectral diffusion.

The main assumption of this model is that in some range of temperature and molecular density, $P$ can become finite and may drive transition in the structural field $m$, which controls long-wavelength density/dielectric fluctuations. 

\paragraph{Mesoscopic structural field $m(\mathbf{r})$.}
$m(\mathbf{r})$ is a coarse-grained scalar field measuring the amplitude of cavity-modified long-wavelength liquid heterogeneity, $m \sim \delta\rho_{\rm cg}(\mathbf{r})$ (coarse-grained density fluctuation).
It is not a microscopic structural variable ({\it i.e.}, the model does not attempt to resolve molecular-scale packing, first-shell correlations, or a specific cluster geometry) but a slow structural/dielectric mode of the liquid whose renormalization enhances the static structure factor at $q \to 0$ (long-wavelength limit). 
In the normal liquid $m = 0$. A nonzero $m$ signals a VSC-modified mesoscopic state with enhanced correlation length and strongly enhanced Rayleigh scattering. 

The physical hierarchy is therefore explicit: the state of the cavity determines $P$, and $m$ is a \emph{secondary} structural field whose long-wavelength susceptibility is renormalized by the collective vibrational polarization intensity $P^2$: as $P^2$ grows, the effective structural stiffness decreases, and the structural field can develop as a secondary instability once $P^2$ is sufficiently large. See details below in the free-energy discussion. 
This implies a probe hierarchy: IR and polaritonic spectroscopies couple to $P$ and are sensitive to the first (vibrational) ordering step; Rayleigh scattering couples to the mesoscopic structural response and becomes large only after the structural spinodal is reached. See details in the Rayleigh-scattering discussion. 

\subsection{Free energy and phase transition} \label{sec:GL}
The simplest GL free-energy functional defined by such two fields, consistent with $P \to -P$ and $m \to -m$ symmetry, and capable of showing the desired transition is
\begin{equation}
  \begin{aligned}
  F[m,P] = \int d^3r\,\Bigl[
    &\frac{a_m}{2}m^2 + \frac{b_m}{4}m^4
      + \frac{\kappa}{2}(\nabla m)^2 \\
    &+ \frac{a_P}{2}P^2 + \frac{b_P}{4}P^4 - \frac{g}{2}\,m^2P^2
  \Bigr],
  \end{aligned}
  \label{eq:free_energy}
\end{equation}
in which the parameters $a_m$, $b_m$, $\kappa$, $a_P$, $b_P$, and $g$ control the system behavior. 
Physically, $a_P$ controls the transition in the macroscopic polarization while $a^\mathrm{eff}_m = a_m - gP^2$ (see Eq.~\ref{eq:aeff}) is a renormalized thermodynamic structural mass which decreases as the collective vibrational polarization grows -- a mesoscopic electrostrictive-like mass-renormalization effect, proportional to the polarization \emph{intensity} $P^2$.
In Eq.~\ref{eq:free_energy}, we treat $P$ as a spatially uniform collective vibrational polarization corresponding to the cavity-coupled bright mode ($k_\parallel = 0$). Therefore, no gradient term for $P$ is included. Spatial fluctuations are instead captured by the structural order parameter $m(\bf r)$, which encodes mesoscopic density/dielectric heterogeneity. 
Potential extension of the order parameters and alternative forms of the GL functional are discussed in Sec.~I of the Supporting Information. 

We further assume that the control parameter $a_P$ depends on several physical variables according to
\begin{equation}
  a_P(\Delta,\Omega,T) = \alpha_T(T - T_0) + \alpha_\Delta\,\Delta^2
  - \alpha_\Omega\,\Omega^2,
  \label{eq:aP}
\end{equation}
where $T$, $\Delta$, and $\Omega$ are temperature, detuning, and collective Rabi splitting, respectively; $\alpha_T$, $\alpha_\Delta$, and $\alpha_\Omega$ are positive coefficients. 
The constant $T_0$ should be understood as a phenomenological intercept of the coarse-grained GL mass, not as a physical transition temperature of the uncoupled liquid. 
Equivalently, one may write $a_P=a_{P,0}+\alpha_T T+\alpha_\Delta\Delta^2-\alpha_\Omega\Omega^2$, with $a_{P,0}=-\alpha_T T_0$. 
In the numerical calibration below we use $T_0=5^\circ$C to set the baseline offset of this local phenomenological parametrization.
This form expresses the expectation that temperature and detuning suppress $P$, the latter encodes resonance selectivity in a sign-independent form; while collective light-matter coupling promotes $P$ through $\Omega^2$ according to symmetry.

The free energy form adopted by Eqs.~\ref{eq:free_energy} and \ref{eq:aP} imposes a transition from the normal liquid ($a_P > 0$) to a state with a finite coarse-grained vibrational order amplitude, represented phenomenologically by $P^2$ ($P \neq 0$) at $a_P<0$. 
$a_P = 0$ defines the primary vibrational GL phase boundary at temperature
\begin{equation}
  T_c(\Omega,\Delta) = T_0 + \frac{\alpha_\Omega\Omega^2 - \alpha_\Delta\Delta^2}
  {\alpha_T}.
  \label{eq:Tc}
\end{equation}
Details of the saddle-point analysis and order-parameter solutions are provided in Sec.~II of the Supporting Information. Eq.~\ref{eq:Tc} implies that at resonance ($\Delta = 0$), $T_c$ scales linearly with $\Omega^2$.

Inside the ordered phase, the coupling $-\dfrac{g}{2}m^2P^2$ renormalizes the quadratic stiffness of the structural mode. Evaluating the Hessian of the free energy (Eq.~\ref{eq:free_energy}) at $m=0$ gives the effective inverse structural susceptibility:
\begin{align}
  a_m^{\rm eff}&\equiv
  \frac{\partial^2 f}{\partial m^2}\bigg|_{m=0}
  =a_m-gP^2 \notag\\
  &=a_m - \frac{g}{b_P}\bigl[
        \alpha_\Omega\Omega^2 - \alpha_\Delta\Delta^2 - \alpha_T(T-T_0)
      \bigr],
  \label{eq:aeff}
\end{align}
valid for $a_P<0$. In the second line of Eq.~\ref{eq:aeff}, we have used the $P$-only branch solution $P^2=-a_P/b_P$ (see derivation in Sec.~II of the Supporting Information) to give the explicit dependence on $\Delta$, $\Omega$, and $T$. For $a_P\geq0$, $a_m^{\rm eff}=a_m$. Eq.~\ref{eq:aeff} is the curvature of the $m$-potential at $m=0$ and controls the long-wavelength structural susceptibility. 

Two distinct boundaries emerge from this construction.
(i)~The \emph{primary vibrational GL instability} $a_P=0$: the onset of collective vibrational polarization ($P\neq0$), defining $T_c$ (Eq.~\ref{eq:Tc}) and the first GL phase boundary.
(ii)~The \emph{structural spinodal} $a_m^{\rm eff}=0$ ($gP^2=a_m$): a secondary instability lying strictly inside the ordered-$P$ region ($a_P<0$), at which the $m=0$ branch loses stability and structural order can emerge with $m^2=(gP^2-a_m)/b_m$.
The phase diagram separated by the two boundaries above is discussed later in the phase-diagram discussion. 

Fig.~\ref{fig:concept}b summarizes the structural part of this GL picture.
For $a_m^{\rm eff}>0$, the $m=0$ state is locally stable and the liquid remains structurally homogeneous at the mesoscopic level.
As $P^2$ grows inside the ordered-$P$ region, $a_m^{\rm eff}=a_m-gP^2$ decreases and the structural Landau landscape becomes progressively flatter.
At the structural spinodal $a_m^{\rm eff}=0$, the $m=0$ branch loses stability; for $a_m^{\rm eff}<0$, the Landau minimum shifts to nonzero $m$, signaling spontaneous mesoscopic structural ordering.

\subsection{Rayleigh scattering intensity} \label{sec:Rayleigh}
The final assumption of our model concerns the way by which the observed Rayleigh scattering signal is associated with the structural field $m(\bf r)$, that was assumed to control long-wavelength density/dielectric heterogeneity. 
Note that in the proposed model, the collective vibrational polarization $P$ does not directly control the Rayleigh signal; rather, it softens the structural field $m(\bf r)$ through the effective mass renormalization in Eq.~\ref{eq:aeff}, thereby enhancing the long-wavelength susceptibility of $m(\bf r)$.
The experimentally measured Rayleigh enhancement ratio, $I_c / I_u$ where $I_c$ is the Rayleigh intensity measured under the VSC condition and $I_u$ is the uncoupled or off-resonant Rayleigh intensity measured under the same density and temperature, is assumed to be described by 
\begin{align} \label{eq:readout}
    I_c / I_u - 1 = I_0\, S_R(a_m^{\rm eff}),
\end{align}
where $I_0$ is an experimental calibration parameter that describes the maximal excess contrast, and $S_R(a_m^{\rm eff})$ is a sigmoid function of $a_m^{\rm eff}$, defined as
\begin{equation}
  S_R(a_m^{\rm eff}) =
  \left[
  1+\exp\left(
  \frac{a_m^{\rm eff}}{\sigma_a}
  \right)
  \right]^{-1},
  \label{eq:sigmoid}
\end{equation}
where $\sigma_a$ is a phenomenological width representing inhomogeneous broadening of the local structural-softening condition. By construction, $S_R$ saturates toward $1$ deep in the softened region ($a_m^{\rm eff}\ll0$), decays toward $0$ on the stable side ($a_m^{\rm eff}\gg0$), and crosses one-half exactly at $a_m^{\rm eff}=0$, i.e.\ at the structural spinodal.
Ideally, we expect a Heaviside function $\Theta(-a_m^{\rm eff})$ for Rayleigh enhancement: in the ideal spatially uniform, infinite-resolution limit, Rayleigh visibility would switch on discontinuously at the structural spinodal, whereas finite spatial/temporal averaging and cavity inhomogeneous broadening smooth this ideal step into the sigmoid form of Eq.~\ref{eq:sigmoid}. More details are provided in Sec.~III-B of the Supporting Information.

To conclude this section, we emphasize that this GL model relies on a free energy functional constructed in accord with experimental observations as shown in the following section, without providing a theoretical basis for its proposed structure. Importantly, we suggest testable predictions that follow from this model, which can be used to validate it and motivate further studies into its microscopic origin.

\section{Calibration and Comparison with Existing Observations}
In this section, we show that with a proper calibration of the GL parameters, our proposed model can capture the experimental observations of Ref.~\citenum{Sandeep2024}. 
In order to minimize the number of independent parameters, we take
$T_0 = 5.0\,^\circ$C and $a_m=b_m=\kappa=b_P=1$, as listed in Table~\ref{tab:gl_params}; these choices fix the normalization of the structural and vibrational fields and the dimensionless length scale.

\subsection{Model parameters} \label{sec:parameters}
Without loss of generality, we focus on toluene--benzene-D6 binary solvent mixtures with the C--H stretching vibration of toluene at 3027 cm$^{-1}$ under VSC, as studied in Ref.~\citenum{Sandeep2024}. 
Parameters are separated into two tables reflecting their distinct roles.
Table~\ref{tab:gl_params} contains the GL coefficients that define the thermodynamic model, which are reported in field-normalized units, with $\Omega$ and $\Delta$ expressed in cm$^{-1}$.
The experimentally extractable combinations are $\alpha_\Delta/\alpha_\Omega=2.50$, which
controls the detuning curvature of the threshold, and
$\alpha_T/\alpha_\Omega=8.58\,\mathrm{cm}^{-2}\,^\circ\mathrm{C}^{-1}$, which gives the
threshold shift in $\Omega^2$ per degree Celsius. Equivalently,
$dT_c/d\Omega^2=\alpha_\Omega/\alpha_T=0.117\,^\circ\mathrm{C\,cm}^{2}$ controls the critical temperature scaling with $\Omega^2$.
The coupling $g\approx0.233$ is chosen accordingly so that the structural spinodal $a_m^{\rm eff}=0$ falls around $\Omega^2 \simeq 1000$ cm$^{-2}$ at $T = 0$ $^\circ$C.
The calibrated coupling satisfies $b_m b_P>g^2$, so the quartic part of the minimal two-field GL free energy remains bounded with $m^2$-$P^2$ coupling.

\begin{table}[h]
  \caption{GL model parameters.}
  \label{tab:gl_params}
  \small
  \begin{tabular}{lll}
    Symbol & Value & Physical role \\
    \hline
    $T_0$         & $5.0\,^\circ$C             & temperature intercept \\
    $\alpha_T$    & $3.43\times10^{-2}$        & thermal suppression \\
    $\alpha_\Delta$ & $1.0\times10^{-2}$       & detuning penalty \\
    $\alpha_\Omega$ & $4.0\times10^{-3}$       & collective softening \\
    $a_m$         & $1.0$                      & bare inverse susceptibility \\
    $g$           & $0.233$                    & $P^2$--$m^2$ coupling \\
    $b_P$         & $1.0$                      & quartic stabilization of $P$ \\
    $b_m$         & $1.0$                      & quartic stabilization of $m$ \\
    $\kappa$      & $1.0$                      & gradient stiffness \\
  \end{tabular}
\end{table}

Table~\ref{tab:observable_params} contains the additional parameters used to convert the stability of the
structural field to Rayleigh observable: $\sigma_a$ is a phenomenological width representing spatial, temporal, and cavity
inhomogeneous broadening of the local structural-softening condition, while $I_0$ is an
experimental Rayleigh contrast calibration parameter for the absolute, unnormalized
signal. Since the plotted Rayleigh curves are normalized excess signals, the numerical
value of $I_0$ is not used in generating the figure data. 
$\sigma_a$ and $I_0$ enter only through
Eqs.~\ref{eq:readout}-\ref{eq:sigmoid}, not through the GL free energy and do not alter the instability threshold or any analytic scaling law.
Additional discussion of the parameter sensitivity is provided in Sec.~III-A of the Supporting Information.

\begin{table}[h]
  \caption{Other parameters (measurement functional only).}
  \label{tab:observable_params}
  \small
  \begin{tabular}{lll}
    Symbol          & Value                      & Physical role \\
    \hline
    $\sigma_a$      & $0.0215$                   & Sigmoid width (mass units) \\
    $I_0$           & N/A                         & absolute Rayleigh contrast \\
  \end{tabular}
\end{table}

\subsection{Phase diagram} \label{sec:phase-diagram}
The GL model encodes a phase diagram that is organized by two nested boundaries, as discussed above in the free-energy discussion. 
The first is the primary vibrational GL instability, $a_P=0$, where the normal liquid gives way to a vibrationally polarized regime with $P\neq 0$. This regime should already display IR/polaritonic signatures. 
As $P^2$ grows further, the structural mass $a_m^{\rm eff}=a_m-gP^2$ is reduced; when $a_m^{\rm eff}=0$, the system reaches the secondary structural spinodal, beyond which the structurally ordered regime and strong Rayleigh enhancement emerge.

\begin{figure}[htbp]
  \centering
  \includegraphics[width=\columnwidth]{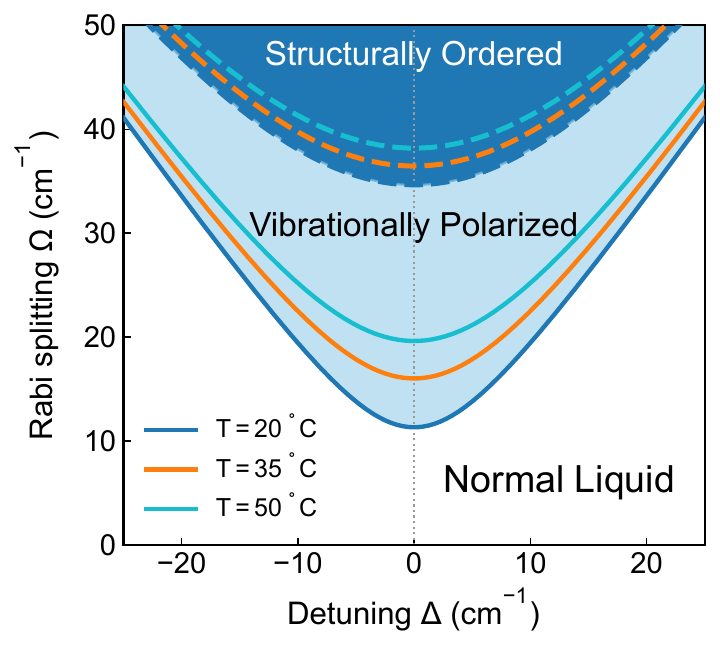}
  \caption{%
    \textbf{Phase diagram in the $(\Delta,\Omega)$ plane.}
    The temperatures are $T = 20$, $35$, $50\,^\circ$C.
    Solid curves: primary vibrational GL boundary $a_P = 0$.
    Dashed curves: structural spinodal $a_m^{\rm eff}=0$, equivalently $gP^2=a_m$ (Eq.~\ref{eq:aeff}).
    Below the solid curve lies the normal liquid; between the solid and dashed curves lies the vibrationally polarized regime, where IR/polaritonic signatures are present but Rayleigh enhancement remains weak; beyond the dashed curve lies the structurally ordered regime where the Rayleigh enhancement becomes strong.
    Raising temperature shifts both boundaries upward, shrinking the ordered window.
  }
  \label{fig:phase_diagrams}
\end{figure}

Using the parameters given in Table~\ref{tab:gl_params}, Fig.~\ref{fig:phase_diagrams} shows the phase diagram in the $(\Delta, \Omega)$ plane at different temperatures.
The solid curves trace the boundary of the primary vibrational polarization GL instability $a_P=0$, and the dashed curves trace the structural spinodal $a_m^{\rm eff} = 0$ (Eq.~\ref{eq:aeff}) or equivalently the equality $a_m = g P^2$.
Below the solid curve is the \emph{normal liquid regime}, in which collective vibrational polarization is absent and structural and spectroscopic probes remain at their uncoupled baselines.
Between the solid and dashed curves lies the \emph{vibrationally polarized regime}: the primary vibrational order has turned on and the structural susceptibility is enhanced, but mesoscopic structural order is absent, remaining on the stable side of the structural spinodal ($a_m^{\rm eff} > 0$) and the sigmoid function $S_R(a_m^{\rm eff})$ (Eq.~\ref{eq:sigmoid}) stays very small. In this regime, IR spectroscopy should already show mode splitting and modified linewidth, while the Rayleigh enhancement remains weak.
Beyond the dashed structural spinodal lies the unstable side of the structural spinodal ($a_m^{\rm eff} < 0$), which is the \emph{structurally ordered regime} with strong Rayleigh enhancement, as the function $S_R(a_m^{\rm eff})$ approaches 1.
This illustrates a two-step hierarchy: IR/polaritonic spectral changes can precede strong Rayleigh enhancement, which appears only near or after the structural spinodal.

\subsection{Thermal collapse of the Rayleigh enhancement}
The GL model can account for the observed thermal collapse of the Rayleigh enhancement because increasing temperature raises $a_P$ (see Eq.~\ref{eq:aP}), reduces $P^2 = -a_P / b_P$, and weakens the structural response entering the Rayleigh signal $S_R(a_m^{\rm eff})$. Thus a scan in $\Omega^2$ at fixed temperature and a scan in temperature at fixed $\Omega$ probe the same control parameter from opposite directions: stronger collective coupling moves the system into the structurally ordered regime, while heating brings it back to the unstructured side of the spinodal.

\begin{figure*}[htbp]
  \centering
  \includegraphics[width=0.95\textwidth]{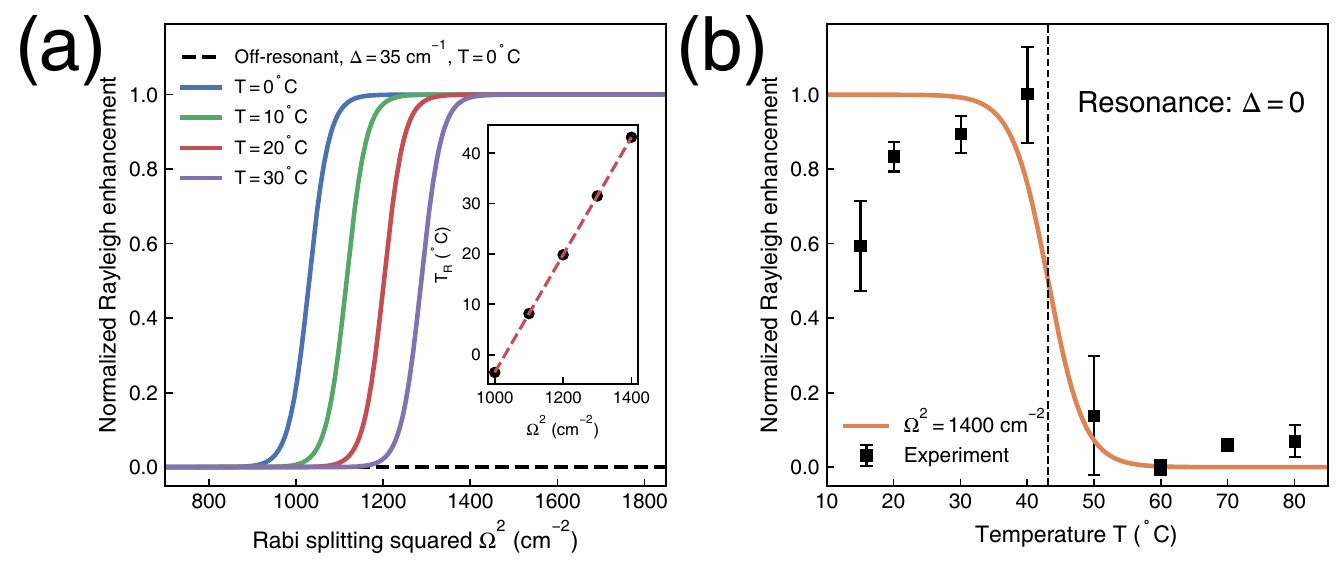}
  \caption{%
    \textbf{Collective coupling threshold and thermal collapse of the Rayleigh enhancement.}
    The Rayleigh enhancement ratios are calculated from Eqs.~\ref{eq:readout}-\ref{eq:sigmoid} with intensities normalized.
    (a) Normalized Rayleigh enhancement vs Rabi splitting squared $\Omega^2$ at
    $T = 0$, $10$, $20$, $30\,^\circ$C (on resonance, $\Delta=0$) and
    an off-resonance case at detuning $\Delta = 35\,\mathrm{cm}^{-1}$, $T = 0\,^\circ$C
    (dashed).
    The off-resonance threshold lies outside the plotted window, so the signal stays at background.
    Inset: Rayleigh collapse temperature $T_R$, defined by $a_m^{\rm eff}(T_R)=0$, vs $\Omega^2$ (circles) with linear fit (red dashed).
    (b) Normalized Rayleigh enhancement vs temperature at
    $\Omega^2 = 1400\,\mathrm{cm}^{-2}$, $\Delta = 0$.
    The sharp collapse near $T_R$ reflects the temperature dependence of the structural
    response entering the sigmoid Rayleigh signal $S_R(a_m^{\rm eff})$.
    Black squares with vertical error bars are experimental data digitized from Ref.~\citenum{Sandeep2024}, shown for visual comparison.
  }
  \label{fig:temperature}
\end{figure*}

Fig.~\ref{fig:temperature}a shows the coupling dependence expressed in terms of $\Omega^2$ at several temperatures.
The plotted curves are computed from Eqs.~\ref{eq:readout}-\ref{eq:sigmoid}; because the Rayleigh excess intensity is normalized, the absolute
amplitude $I_0$ in Eq.~\ref{eq:readout} cancels out.
On resonance, the threshold of the Rayleigh enhancement shifts upward by $\approx 86\,\mathrm{cm}^{-2}$ per $10\,^\circ$C when temperature increases, consistent with the slope $\alpha_T/\alpha_\Omega$ (fitted parameters discussed above); 
while the off-resonance case remains at background.
The inset shows the Rayleigh collapse temperature $T_R$ against $\Omega^2$.
Here $T_R$ is defined as the structural-spinodal temperature satisfying $a_m^{\rm eff}(T_R)=0$ (equivalently $P^2=a_m/g$), and is distinct from the primary GL instability temperature $T_c$: $T_R = T_c - b_P a_m/(g\alpha_T)$ at fixed Rabi splitting and detuning. Because the sigmoid function
$S_R(a_m^{\rm eff})$ is centered at $a_m^{\rm eff}=0$, $T_R$ also marks the midpoint of the Rayleigh collapse in the model.
At fixed detuning it scales linearly with $\Omega^2$, in agreement with the experiment~\cite{Sandeep2024}.

Fig.~\ref{fig:temperature}b further shows the same construction as a sharp thermal collapse of Rayleigh enhancement at $T_R \sim 43\,^\circ$C for fixed $\Omega^2 = 1400\,\mathrm{cm}^{-2}$ and $\Delta = 0$. 
Experimental data digitized from Sandeep \textit{et al.}~\cite{Sandeep2024} (black squares with vertical error bars) are shown for visual comparison. 
The model is calibrated to the threshold location and collapse trend rather than fitted to the full temperature dependence of the normalized experimental signal.

\section{Predicted Consequences and Proposed Experiments}
Having demonstrated that the constructed framework can capture existing observations, we turn to examining its consequences regarding yet unobserved phenomenology. 
We outline below several signatures in the mesoscopic structural and dynamical responses, suggesting specific experimental handles that may be used to test or potentially falsify the proposed mechanism.

\subsection{Pair correlation length} \label{sec:pair_correlation}
The thermodynamic effective mass $a_m^{\rm eff}$ controls the long-wavelength structural susceptibility and determines the corresponding spatial correlation length. 
The GL model therefore predicts a structural signature beyond optical scattering. By the proposed model, VSC could enhance long-range density correlations while leaving molecular-scale structure essentially unchanged.
With the gradient term in Eq.~\ref{eq:free_energy}, the Gaussian structural fluctuation spectrum takes the Ornstein--Zernike (OZ) form~\cite{Ornstein_1914, Chaikin_Lubensky_1995, Hansen_McDonald_2013},
\begin{equation}
  \begin{aligned}
  S(q)&=\frac{k_\mathrm{B}T}{a_m^{\rm eff}+\kappa q^2}.
  \end{aligned}
  \label{eq:Sq}
\end{equation}
See details in Sec.~IV of the Supporting Information.
Here $S(q)$ denotes the coarse-grained long-wavelength structure factor associated with $m(\bf r)$, which implies a corresponding long-distance contribution to the molecular total correlation function $h(r)\equiv g(r)-1$,
with the bare structural mass $a_m$ replaced by the VSC-renormalized mass $a_m^{\rm eff}$ inside the cavity. 
The long-wavelength limit is
\begin{align} \label{Sq-lw}
    S(q\to0)=\frac{k_\mathrm{B}T}{a_m^{\rm eff}}.
\end{align}
We also define $S_{\rm out}(0) = k_\mathrm{B}T / a_m$, which represents the uncoupled counterpart of Eq.~\ref{Sq-lw}. Note that in Eqs.~\ref{eq:Sq}-\ref{eq:sqw}, $T$ denotes the absolute temperature, whereas the temperature differences entering Eq.~\ref{eq:aP} may equivalently be expressed in degrees Celsius.

The corresponding long-distance pair-correlation tail is
\begin{equation}
  h(r)\equiv g(r)-1 \sim A\,\frac{e^{-r/\zeta}}{r},
  \label{eq:gr_tail}
\end{equation}
where $\zeta$ determines the correlation length,
\begin{align}
    \zeta&=\sqrt{\frac{\kappa}{a_m^{\rm eff}}}.
\end{align}
Eq.~\ref{eq:gr_tail} is the real-space inverse-Fourier counterpart of Eq.~\ref{eq:Sq}. 
We note that Eq.~\ref{eq:gr_tail} does not describe molecular-scale packing or the short-range oscillatory part of $g(r)$.
These quantitative OZ expressions Eqs.~\ref{eq:Sq} and~\ref{eq:gr_tail} are used only on the stable side, $a_m^{\rm eff}>0$.
With the calibrated coupling $g=0.233$, the structural spinodal occurs at $P^2=a_m/g\approx4.29$.

\begin{figure*}[htbp]
  \centering
  \includegraphics[width=0.95\textwidth]{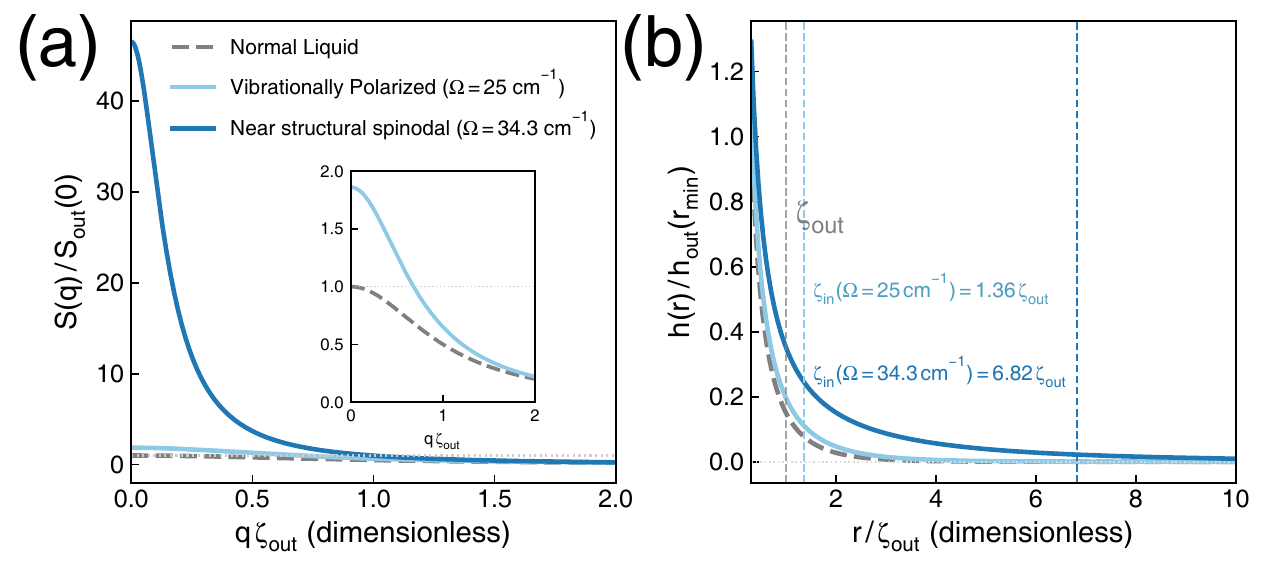}
  \caption{%
    \textbf{Implications of the GL model for the structure factor and the pair correlations.}
    (a) OZ response
    $S(q) = k_\mathrm{B}T/(a_m^{\rm eff}+\kappa q^2)$
    normalized to $S_{\rm out}(0)$, valid for $a_m^{\rm eff}>0$.
    Gray dashed: outside cavity ($a_m^{\rm eff} = a_m = 1.0$).
    Light blue: vibrationally polarized regime at $\Omega = 25\,\mathrm{cm}^{-1}$, $\Delta = 0$,
    $T = 20\,^\circ$C, with weak low-$q$ enhancement
    ($S(0)/S_{\rm out}(0) \approx 1.86$).
    Dark blue: stable-side representative approaching the structural spinodal at
    $\Omega \approx 34.3\,\mathrm{cm}^{-1}$, $\Delta = 0$, $T = 20\,^\circ$C
    ($a_m^{\rm eff}=\sigma_a\approx0.0215$, just above the spinodal).
    Low-$q$ enhancement: $S(0)/S_{\rm out}(0) \approx 46.5$.
    Inset: low-$q$ view of the outside-cavity and vibrationally polarized curves.
    (b) Real-space OZ tail $h(r)\equiv g(r)-1 \propto e^{-r/\zeta}/r$, the inverse-Fourier counterpart of the OZ line shapes in panel (a), plotted from
    $r_{\rm min} = 0.3\,\zeta_{\rm out}$ (short-range cutoff).
    The y-axis is normalized by $h_{\rm out}(r_{\rm min})$, where $h_{\rm out}(r)$ denotes the outside-cavity total correlation function.
    Dotted verticals: $\zeta_{\rm out}$ (gray), the vibrationally polarized correlation length
    $\zeta_{\rm in}(\Omega = 25\,\mathrm{cm}^{-1}) \approx 1.36\,\zeta_{\rm out}$ (light blue), and
    $\zeta_{\rm in}(\Omega \approx 34.3\,\mathrm{cm}^{-1}) \approx 6.82\,\zeta_{\rm out}$ (dark blue).
  }
  \label{fig:sq_gr}
\end{figure*}

Figure~\ref{fig:sq_gr} presents the OZ structure factor and pair-correlation tail with representative $a_m^{\rm eff}>0$.
Fig.~\ref{fig:sq_gr}a shows $S(q) = k_\mathrm{B}T/(a_m^{\rm eff}+\kappa q^2)$ (normalized by $S_{\rm out}(0)$) for the normal liquid (gray dashed curve), the vibrationally polarized regime (light blue solid curve), and a stable-side representative near the structural spinodal (dark blue solid curve).
For the vibrationally polarized case ($\Omega = 25\,\mathrm{cm}^{-1}$, $\Delta = 0$, $T=20\,^\circ$C), the low-$q$ response is enhanced only weakly, $S(0)/S_{\rm out}(0)\approx1.86$ and $\zeta_{\rm in}(\Omega = 25\,\mathrm{cm}^{-1})/\zeta_{\rm out}\approx1.36$.
For the near-spinodal case, $\Omega\simeq34.3\,\mathrm{cm}^{-1}$ is chosen so that $a_m^{\rm eff}=\sigma_a\approx0.0215$, a representative positive mass just above the structural spinodal, giving $S(0)/S_{\rm out}(0)\approx46.5$ and $\zeta_{\rm in}(\Omega \simeq 34.3\,\mathrm{cm}^{-1})/\zeta_{\rm out}\approx6.82$.

Fig.~\ref{fig:sq_gr}b shows the corresponding real-space OZ tail $h(r)\equiv g(r)-1 \approx A\,e^{-r/\zeta}/r$, obtained as the inverse-Fourier transform of the same OZ structure factors shown in Fig.~\ref{fig:sq_gr}a.
The vibrationally polarized curve shows only a modest extension of the tail, $\zeta_{\rm in}(\Omega = 25\,\mathrm{cm}^{-1})\approx1.36\,\zeta_{\rm out}$, whereas the near-spinodal tail decays over $\zeta_{\rm in}(\Omega \simeq 34.3\,\mathrm{cm}^{-1}) \approx 6.82\,\zeta_{\rm out}$, producing the slow-decaying structural correlations responsible for the enhanced Rayleigh signal at small $q$.
If $\zeta_{\rm out} \sim 1$--$2$~nm (typical for molecular liquids), the predicted ratio $\zeta_{\rm in} /\zeta_{\rm out} \sim 6.82$ would correspond to $\zeta_{\rm in}(\Omega \simeq 34.3\,\mathrm{cm}^{-1}) \sim 7$--$14$~nm --- a length scale potentially testable by small-angle X-ray or neutron scattering (SAXS/SANS) and dynamic light scattering (DLS) measurements.

For larger Rabi splitting squared $\Omega^2$ that gives $a_m^{\rm eff}<0$, the $m=0$ Gaussian OZ expression is no longer the correct thermodynamic result. That regime would require expansion about the nonzero-$m$ branch. Fig.~\ref{fig:sq_gr} therefore describes the evolution of the system structure as we approach the spinodal from the stable side.

\subsection{Dynamic structure factor and low-frequency response}

The static OZ response discussed above has a direct dynamical counterpart. To describe the slow structural/dielectric fluctuations represented by $m({\bf r})$, we supplement the quadratic GL functional with an overdamped time-dependent GL equation~\cite{HohenbergHalperin1977} for each Fourier component $m_{\bf q}(t)$. On the stable side of the structural spinodal, define
\[
A_q \equiv a_m^{\rm eff}+\kappa q^2 .
\]
The expressions below are predictions of the thermodynamic mass and are valid for $a_m^{\rm eff}>0$; they are distinct from the phenomenological Rayleigh intensity $I_c/I_u$, which is instead obtained from $a_m^{\rm eff}$ through the sigmoid function $S_R$ (Eq.~\ref{eq:sigmoid}).

The minimal overdamped dynamics is
\begin{equation}
\partial_t m_{\bf q}(t)
=
-\Gamma_m A_q\,m_{\bf q}(t)
+
\eta_{\bf q}(t),
\end{equation}
with thermal noise
\begin{equation}
\langle \eta_{\bf q}(t)\eta_{-\bf q}(t')\rangle
=
2\Gamma_m k_\mathrm{B}T\,\delta(t-t').
\end{equation}
Here $\Gamma_m$ is a phenomenological kinetic coefficient for the slow structural/dielectric coordinate. This choice gives the correct equal-time fluctuation,
\[
S(q)
=
\langle |m_{\bf q}|^2\rangle
=
\frac{k_\mathrm{B}T}{A_q},
\]
recovering the static OZ form of Eq.~\ref{eq:Sq}. Further derivation is provided in Sec.~IV of the Supporting Information.

The corresponding retarded susceptibility is
\begin{equation}
\chi(q,\omega)
=
\frac{1}{A_q-i\omega/\Gamma_m},
\end{equation}
and the classical fluctuation--dissipation relation~\cite{Nitzan} gives the dynamic structure factor
\begin{align} \label{eq:dynamic_struc}
S(q,\omega)
&=
\int_{-\infty}^{\infty}dt\,e^{i\omega t}
\langle m_{\bf q}(t)m_{-\bf q}(0)\rangle \notag\\
&=
\frac{2\Gamma_m k_\mathrm{B}T}
{\omega^2+\Gamma_m^2 A_q^2}.
\end{align}
Equivalently,
\begin{equation} \label{eq:sqw}
S(q,\omega)
=
\frac{2\tau_q S(q)}
{1+\omega^2\tau_q^2},
\qquad
\tau_q^{-1}=\Gamma_m A_q .
\end{equation}
Thus, reducing $a_m^{\rm eff}$ toward the structural spinodal enhances the static low-$q$ structure factor and, at the same time, slows the corresponding structural dynamics. At fixed small $q$, the dynamic structure factor $S(q,\omega)$ is therefore predicted to develop three correlated signatures: larger integrated low-frequency intensity, a narrower quasielastic central peak, and a longer structural/dielectric relaxation time.

This response should be understood as the quasielastic central-peak signature of the slow structural/dielectric mode represented by $m(\mathbf r)$, rather than the motion of a new vibrational mode. The most direct probes are dynamic light scattering, depolarized light scattering, and quasielastic neutron or X-ray scattering, which measure the low-frequency spectral density and linewidth associated with $S(q,\omega)$. Low-frequency dielectric or THz spectroscopy may provide a complementary probe if this structural mode carries sufficient dielectric weight. A positive test would show samples that show Rayleigh enhancement also exhibit enhanced and slowed low-$q$  (meaning $q \zeta_{\rm out} \ll 1$) structural fluctuations while their high-$q$ (meaning $q \zeta_{\rm out} \gg 1$) liquid structure remains approximately bulk-like. Conversely, the absence of any corresponding low-$q$ quasielastic enhancement or slowing would cast doubt on the proposed GL mechanism.

The results above assume that $m(\mathbf r)$ is a non-conserved
structural/dielectric coordinate. If $m$ is instead identified with a
strictly conserved density field, the dynamics becomes diffusive and
the small-$q$ relaxation rate acquires an additional factor of $q^2$,
while the static OZ response remains unchanged. See details in
Supporting Information, Sec.~IV.

\subsection{Real-space and near-field imaging}
We emphasize, however, neither the experiment in Ref.~\citenum{Sandeep2024} nor the GL model here directly resolve a specific molecular cluster geometry or size distribution. 
Real-space imaging, such as near-field optical imaging, interferometric scattering microscopy, or cryogenic/frozen-snapshot approaches, offers a complementary real-space perspective to scattering techniques (but is a less direct test of the GL model). While the highly dynamic nature of these mesoscopic density fluctuations poses challenges for conventional optical microscopy in simple liquids, VSC experiments performed in less volatile fluid media -- such as polymeric matrices, supramolecular gels, or liquid crystals -- could provide an ideal platform to directly freeze and map these spatially heterogeneous dielectric domains, if such domains exist.

However, the absence of directly visible clusters would not by itself rule out the GL mechanism, because the primary prediction is an enhanced long-wavelength correlation function, not a static cluster morphology. Conversely, any observed spatial heterogeneity should be checked against detuning, temperature, and concentration controls to establish that it follows the VSC resonance rather than ordinary cavity, surface, or drying effects.
\section{Discussion}
The present GL model begins at the mesoscopic level by introducing two experimentally motivated order parameters and constraining their phenomenological coefficients using the observed trends. 
The model involves two coarse-grained fields: $P$, the cavity-selected bright-mode vibrational polarization, and $m$, a scalar structural/dielectric field describing long-wavelength liquid heterogeneity.
The model does not attempt to derive these variables, their coefficients, or the optical Rayleigh signal from a microscopic Hamiltonian. Instead, the Rayleigh signal is treated as a phenomenological measurement functional that maps the structural effective mass $a_m^{\rm eff}$ to the measured optical contrast through $S_R(a_m^{\rm eff})$. The observed Rayleigh ``transition'' should therefore be understood as a finite-resolution optical manifestation of the structural spinodal $a_m^{\rm eff}=0$, rather than as an additional phase boundary introduced independently of the GL theory. In the ideal uniform limit, this reduces to the step function $\Theta(-a_m^{\rm eff})$; finite spatial, temporal, and cavity inhomogeneous broadening round the step into the sigmoid form in Eq.~\ref{eq:sigmoid}, with $\sigma_a$ setting the width and $I_0$ setting the saturated excess Rayleigh contrast, as discussed in Sec.~III-B of the Supporting Information.

With this construction, the GL model captures the central experimental pattern of VSC-enhanced Rayleigh scattering: resonance selectivity, an abrupt collective threshold, and thermal collapse, even though polaritonic spectral signatures can persist when the Rayleigh enhancement is weak. The essential physical point is a two-step hierarchy. The first boundary, $a_P=0$, marks the onset of collective vibrational polarization and therefore of IR/polaritonic signatures. The second boundary, $a_m^{\rm eff}=0$, is the structural spinodal at which the mesoscopic structural field becomes unstable and the Rayleigh signal rapidly turns on. This separation is conceptually consistent with recent numerical work showing that, in energetically disordered molecular ensembles, spectroscopic Rabi splitting does not by itself guarantee polariton delocalization; substantially stronger collective coupling is required before delocalized polaritonic states emerge~\cite{Xiong_2025}. In the present model, an analogous distinction appears between observing IR/polaritonic signatures and observing a strong Rayleigh contrast. 

Although the microscopic origin of the order parameters $P$ and $m(\bf r)$ remains unresolved, their introduction at the mesoscopic level yields a unified and internally consistent description of the observed experimental trends.
The primary field $P$ is probed by IR and polaritonic measurements, including normal-mode splitting, oscillator-strength redistribution, linewidth changes, and multidimensional-IR spectral diffusion. 
These observables characterize the cavity-selected bright vibrational mode, although they do not determine a unique scalar value of $P$. 
The secondary field $m$ should instead be tested through long-wavelength structural and dynamical correlations: Rayleigh scattering measures the optical contrast associated with the structural field, SAXS/SANS can measure the low-$q$ enhancement of $S(q)$ and extract the correlation length $\zeta$, and DLS, depolarized light scattering, or quasielastic neutron/X-ray scattering can test the predicted central peak and structural slowing.
This distinction leads to a direct experimental test of the response hierarchy. At fixed resonance, concentration or temperature sweeps should reveal a vibrationally polarized window in which IR/polaritonic signatures are already present while Rayleigh scattering and low-$q$ structural enhancement remain weak. As the system approaches the structural spinodal, $a_m^{\rm eff}=0$ or equivalently $P^2\simeq a_m/g$ under the present calibration, the Rayleigh signal $S_R(a_m^{\rm eff})$ turns on and should be accompanied by a correlated increase in Rayleigh contrast, $S(q\to0)$, $\zeta$, and the structural relaxation time. 
Conversely, the absence of any corresponding low-$q$ scattering enhancement or slowing in Rayleigh-enhanced samples would strongly constrain the GL mechanism. 
Low-frequency dielectric or THz spectroscopy may provide a complementary probe if the structural mode carries dielectric weight, while spatially resolved Rayleigh imaging should test the predicted broadening of the apparent transition in less uniform cavities or concentration profiles.
A recent experiment by Itatani, {\it et al.} demonstrated that liquid water under vibrational ultrastrong coupling (V-USC) undergoes a collective reorganization of its hydrogen-bond network~\cite{Itatani_chemrxiv2026}.
By decomposing the O--H stretch manifold, they observed an anomalous suppression of thermal disorder driven by extended intermolecular correlations that span across multiple subensembles. This is qualitatively consistent with our prediction of an enlarged low-$q$ correlation length $\zeta$ in Fig.~\ref{fig:sq_gr}b, while respecting the constraint that local, high-$q$ chemical structures remain bulk-like. 

Direct molecular dynamics simulations provide a complementary route to test the microscopic origin of the proposed mesoscopic response while retaining atomistic detail. 
Earlier cavity molecular dynamics simulations of liquid water based on the q-TIP4P/F water model~\cite{Habershon_JCP2009_qTIP4PF} found essentially unchanged O--O radial distribution functions and O--H bond-length distributions under VSC or vibrational ultrastrong coupling~\cite{Li_pnas2020}. 
This result places an important constraint on the present mesoscopic picture: the predicted VSC-associated structural response should appear primarily in the long-wavelength, low-$q$ regime and in the large-$r$ correlation tail, rather than through a substantial rearrangement of local liquid structure.
Future simulations should therefore move beyond local radial distribution functions and directly resolve the low-$q$ structural and dielectric response. 
In practice, this requires computing the small-$q$ limit of the structure factor, $S(q\to0)$, analyzing finite-size scaling in molecular cavities, and tracking dielectric or polarization fluctuations associated with the collective structural mode. 
For water and other strongly polar liquids, an \emph{ab initio}-level description may be important because polarization changes and hydrogen-bond-network rearrangements can be sensitive to the underlying force field and water-model architecture~\cite{Vega_PCCP2011,Cisneros_ChemRev2016,Martelli_JMolLiq2021}. 
Although fully \emph{ab initio} cavity molecular dynamics at the required system sizes is computationally demanding, machine-learning interatomic potentials (MLIPs)~\cite{Behler_PRL2007,Bartok_PRL2010,Zhang_PRL2018_DP,Wang_CPC2018_DeePMD,Zeng_JCP2023_DeePMDkitV2} trained on electronic-structure data, implemented for example with frameworks such as DeePMD-kit~\cite{Zhang_PRL2018_DP,Wang_CPC2018_DeePMD,Zeng_JCP2023_DeePMDkitV2}, may offer a practical route toward \emph{ab initio}-level accuracy beyond empirical fixed-charge water models such as q-TIP4P/F~\cite{Habershon_JCP2009_qTIP4PF}.

Finally, while the present GL model provides a compact phenomenological description of the observed hierarchy, a microscopic derivation remains an essential next step.
A microscopic theory would require deriving the collective bright-mode polarization $P$ and its susceptibility, the coefficients entering $a_P(\Delta,\Omega,T)$, the coupling between $P$ and liquid structural degrees of freedom, and the resulting distribution of local structural-softening conditions from an underlying light--matter Hamiltonian combined with a statistical theory of the liquid. 
Classical density functional theory (cDFT)~\cite{Evans_AdvPhys1979, Lutsko_AdvChemPhys2010, Bui_PRL2025} offers a potential framework for the liquid part of this program: it formulates the equilibrium liquid free energy as a functional of the density field and can, in principle, connect microscopic intermolecular interactions and direct correlation functions to the long-wavelength structural response encoded here by $m(\mathbf r)$. 
In this sense, cDFT provides a microscopic-statistical route toward the coarse-grained liquid of the present GL model. 
It does not, by itself, determine the cavity-selected collective polarization $P$ or its coupling to the structural field $m(\mathbf r)$, which must be derived from the light--matter Hamiltonian. 
Possible starting points for this complementary part include spin-glass-inspired descriptions~\cite{Sidler_ChemRev2026, sidler_arxiv2026}, macroscopic-quantum-state frameworks~\cite{Huo2025}, and Kuramoto-type synchronization models~\cite{Kuramoto1981, Acebron_RMP2005}.

\section{\small $\blacksquare$ ASSOCIATED CONTENT}
{\bf Data Availability Statement}. \\
The data that support the findings of this work are available in \url{https://github.com/Okita0512/VSC-GL-theory}.\\

\noindent {\bf Supporting Information}. \\
The Supporting Information is available free of charge at [url].

The Supporting Information contains discussions of alternative coupling terms; mean-field stationary solutions and stability; parameter identifiability and Rayleigh broadening; and derivations of static and dynamic structural correlations.\\

\section{\small $\blacksquare$ AUTHOR INFORMATION}
Complete contact information is available at: [url]\\

{\bf Notes}\\
The authors declare no competing financial interest.\\

\section{\small $\blacksquare$ ACKNOWLEDGEMENTS}
This work was supported by the European Research Council under ERC-2024-SyG-101167294; UnMySt.
\bibliography{ref}

\end{document}


\title{Supporting Information for\\ A Mesoscopic Ginzburg--Landau Model for Vibrational Strong Coupling Enhanced Rayleigh Scattering in Molecular Liquids}

\author{Wenxiang Ying}
\email{wying3@sas.upenn.edu}
\affiliation{Department of Chemistry, University of Pennsylvania, Philadelphia, Pennsylvania 19104, USA}

\author{Abraham Nitzan}
\email{anitzan@sas.upenn.edu}
\affiliation{Department of Chemistry, University of Pennsylvania, Philadelphia, Pennsylvania 19104, USA}
\affiliation{School of Chemistry, Tel Aviv University, Tel Aviv 69978, Israel}

\maketitle
\tableofcontents

\section{Order-Parameter Assumptions and Alternative Couplings}

\subsection{Minimal free-energy functional}

The minimal free energy is
\begin{equation}
  F[m,P] = \int d^3r \left[
    \frac{a_m}{2}m^2 + \frac{b_m}{4}m^4
    + \frac{\kappa}{2}(\nabla m)^2
    + \frac{a_P}{2}P^2 + \frac{b_P}{4}P^4
    - \frac{g}{2}\,m^2P^2
  \right],
  \label{eq:SI_free_energy}
\end{equation}
with all cavity dependence assigned to the primary vibrational amplitude,
\begin{equation}
  a_P(\Delta,\Omega,T)
  = \alpha_T(T-T_0) + \alpha_\Delta\Delta^2 - \alpha_\Omega\Omega^2 .
  \label{eq:SI_aP}
\end{equation}

In the minimal model $P$ is spatially uniform, and the spatial dependence resides
entirely in $m(\mathbf r)$; the coupling $-\frac{g}{2}m^2P^2$ changes the
curvature in the $m$ direction from $a_m$ to $a_m-gP^2$, and the gradient
coefficient $\kappa$ controls the long-wavelength spatial response of $m$. The
leading cavity correction to $a_P$ is quadratic in $\Omega$ because the free
energy cannot depend on the sign or phase of the microscopic light--matter
coupling.

\subsection{Extensions of the order parameters}

The minimal model treats $P$ as a spatially uniform, real scalar amplitude and
$m$ as Gaussian on the stable side of the instability. Allowing $P(\mathbf r)$
to vary, writing $P(\mathbf r)=|P(\mathbf r)|e^{i\phi(\mathbf r)}$ with an
explicit phase, retaining non-Gaussian fluctuations, or adding higher gradients
and higher-order invariants are possible extensions not required by the
present phenomenology and not analyzed further here.

\subsection{Alternative coupling terms}
\label{sec:closures}

The coupling $-\frac{g}{2}m^2P^2$ gives the effective mass renormalization
\begin{equation}
  a_m^{\rm eff}(P^2)=\frac{\partial^2 F}{\partial m^2}\bigg|_{m=0}=a_m-gP^2
  \label{eq:SI_aeff_closure}
\end{equation}
as an exact consequence of the free energy: this is the curvature of $F$ at
$m=0$, not an independent phenomenological assumption. Eq.~\ref{eq:SI_free_energy}
is symmetric under $m\to-m$ and $P\to-P$. The $P\to-P$ invariance follows from
the phase arbitrariness of the collective vibrational amplitude. The $m\to-m$
invariance, by contrast, is an imposed symmetry of the minimal truncation, not
a generic exact symmetry of a density-like field; relaxing it is considered
below.

\paragraph{Possible origin of the coupling.}
One possible phenomenological origin of $-\frac{g}{2}m^2P^2$ is a weak
dependence of the collective Rabi splitting on the structural field, the
leading form consistent with $m\to-m$ being
\begin{equation}
  \Omega^2(m) = \Omega_0^2 + \frac{\lambda}{2}m^2 + O(m^4).
  \label{eq:SI_Omega_m}
\end{equation}
Substituting into $a_P$ and then into $\frac{a_P}{2}P^2$ gives
$\delta f=-\frac{\alpha_\Omega\lambda}{4}m^2P^2$, equivalent to the minimal
coupling with
\begin{equation}
  g=\frac{\alpha_\Omega\lambda}{2}.
  \label{eq:SI_g_from_lambda}
\end{equation}
This is a symmetry-level interpretation, not a microscopic derivation of $g$.

Relaxing $m\to-m$ admits a source-like term,
\begin{equation}
  F_{\rm ext}[m,P]=\int d^3r\left[
  \frac{1}{2}(a_m-g P^2)m^2
  +\frac{b_m}{4}m^4
  +\frac{\kappa}{2}(\nabla m)^2
  +\frac{a_P}{2}P^2
  +\frac{b_P}{4}P^4
  -h\,mP^2
  \right].
  \label{eq:SI_extended_F}
\end{equation}
For weak $h$, away from $a_m^{\rm eff}=0$, the stationarity equation
$(a_m^{\rm eff}+b_mm^2)m=hP^2$ is solved perturbatively,
\begin{equation}
  m_* = \frac{hP^2}{a_m^{\rm eff}} + O(h^3).
  \label{eq:SI_m_star}
\end{equation}
Substituting back into $\frac12a_m^{\rm eff}m^2-hmP^2$ gives
\begin{equation}
  \delta f(P) = -\frac{h^2P^4}{2\,a_m^{\rm eff}(P^2)} + O(h^4),
  \label{eq:SI_delta_f}
\end{equation}
so that the effective quartic coefficient of the $P$-dependent part of the free
energy is
\begin{equation}
  b_P^{\rm eff} = b_P - \frac{2h^2}{a_m^{\rm eff}(P^2)},
  \label{eq:SI_bPeff}
\end{equation}
not $b_P-h^2/a_m^{\rm eff}$: both the mass term and the source term contribute
at $O(h^2)$ once $m_*$ is substituted, and their sum is twice the magnitude of
either term alone. Because $a_m^{\rm eff}(P^2)=a_m-gP^2$ depends on $P$, this
is a $P$-dependent effective coefficient; expanding through $O(P^4)$,
\begin{equation}
  b_P^{\rm eff} = b_P - \frac{2h^2}{a_m}
  \label{eq:SI_bPeff_quartic}
\end{equation}
at quartic order, with corrections beginning at $O(P^6)$. This perturbative
elimination of $m$ fails as $a_m^{\rm eff}\to0$, where the full stationarity
equation must be solved without expanding in $h$. If $b_P^{\rm eff}<0$,
stabilizing higher-order terms (e.g.\ $m^6$, $P^6$, $m^2P^4$) are required, and
the transition need not remain continuous.

\newpage
\section{Mean-Field Stationary Solutions and Stability}
\label{sec:SI_saddle_point}

\subsection{Stationary equations and the P-only branch}
\label{sec:SI_Ponly_branch}

For uniform fields, Eq.~\ref{eq:SI_free_energy} gives the free-energy density
\begin{equation}
  f(m,P)=
  \frac{a_m}{2}m^2+\frac{b_m}{4}m^4
  +\frac{a_P}{2}P^2+\frac{b_P}{4}P^4-\frac{g}{2}m^2P^2 ,
\end{equation}
with stationarity equations
\begin{align}
  \frac{\partial f}{\partial m}
  &= m\left(a_m+b_m m^2-gP^2\right)=0,
  \label{eq:SI_sp_m}\\
  \frac{\partial f}{\partial P}
  &= \left(a_P+b_P P^2-gm^2\right)P=0.
  \label{eq:SI_sp_P}
\end{align}
Eq.~\ref{eq:SI_sp_m} has two branches, $m=0$ and $m^2=(gP^2-a_m)/b_m$
(structural ordering for $gP^2>a_m$); Eq.~\ref{eq:SI_sp_P} shows that the
primary instability occurs in $P$.

For $a_P>0$ the stable solution is the normal liquid, $P=0$, $m=0$. The
primary instability occurs at $a_P=0$; using Eq.~\ref{eq:SI_aP}, this gives the
critical Rabi splitting
\begin{equation}
  \Omega_c^2(\Delta,T)
  = \frac{\alpha_T(T-T_0)+\alpha_\Delta\Delta^2}{\alpha_\Omega}.
  \label{eq:SI_OmegaC}
\end{equation}

For $a_P<0$, the $m=0$ branch of Eq.~\ref{eq:SI_sp_m} remains stationary, and
the nontrivial branch of Eq.~\ref{eq:SI_sp_P} gives the P-only solution,
\begin{equation}
  P^2 = -\frac{a_P}{b_P},
  \label{eq:SI_nontrivial_branch}
\end{equation}
so that together
\begin{equation}
  P^2=
  \begin{cases}
    0, & a_P\geq0,\\[2pt]
    -a_P/b_P, & a_P<0,
  \end{cases}
  \label{eq:SI_P2_piecewise}
\end{equation}
defines the P-only branch on both sides of the primary instability, with
reduced free energy
\begin{equation}
  f_{\rm red}(P)
  = \frac{a_P}{2}P^2
  +\frac{b_P}{4}P^4.
  \label{eq:SI_fred}
\end{equation}
The structural mean field vanishes on this branch, $\langle m\rangle=0$, with
curvature in the $m$ direction given by Eq.~\ref{eq:SI_aeff_closure}. The
P-only branch is not the complete self-consistent solution once
$a_m^{\rm eff}<0$; the fully coupled branch is derived next.

\subsection{Fully coupled ordered branch}
\label{sec:SI_fully_coupled}

Beyond the point where the P-only branch loses stability, both order
parameters are nonzero and Eqs.~\ref{eq:SI_sp_m}--\ref{eq:SI_sp_P} must be
solved simultaneously, since the P-only relation $P^2=-a_P/b_P$ no longer
holds once $m\neq0$:
\begin{align}
  m^2&=\frac{gP^2-a_m}{b_m},
  \label{eq:SI_m2_of_P2}\\
  P^2&=\frac{gm^2-a_P}{b_P}.
  \label{eq:SI_P2_of_m2}
\end{align}
Solving this linear system for $(m^2,P^2)$ and defining
\begin{equation}
  D \equiv b_m b_P - g^2,
  \label{eq:SI_D_def}
\end{equation}
gives
\begin{align}
  m_0^2 &= -\frac{b_P a_m + g a_P}{D},
  \label{eq:SI_m0_full}\\
  P_0^2 &= -\frac{b_m a_P + g a_m}{D}.
  \label{eq:SI_P0_full}
\end{align}
The quartic form $\frac{b_m}{4}m^4+\frac{b_P}{4}P^4-\frac{g}{2}m^2P^2$ is
bounded from below if and only if
\begin{equation}
  b_m>0,\qquad b_P>0,\qquad D=b_mb_P-g^2>0,
  \label{eq:SI_D_positive}
\end{equation}
and the fully coupled branch exists as a physical solution only when
$m_0^2\geq0$ and $P_0^2\geq0$; otherwise the stationary point is unphysical
and the relevant solution reverts to the P-only or normal branch.

\subsection{Effective free energy and free-energy comparison}
\label{sec:SI_feff}

Eliminating $P$ at fixed $m$ through $P^2(m)=(gm^2-a_P)/b_P$ and substituting
into $f(m,P)$ gives
\begin{equation}
  f_{\rm eff}(m)
  = -\frac{a_P^2}{4b_P}
  + \frac12 a_m^{\rm eff}m^2
  + \frac14 b_m^{\rm eff}m^4,
  \label{eq:SI_feff_m}
\end{equation}
with
\begin{equation}
  a_m^{\rm eff} \equiv a_m+\frac{g\,a_P}{b_P},
  \qquad
  b_m^{\rm eff} \equiv b_m-\frac{g^2}{b_P} = \frac{D}{b_P},
  \label{eq:SI_ameff_via_aP}
\end{equation}
where $a_m^{\rm eff}$ coincides with Eq.~\ref{eq:SI_aeff_closure} evaluated on
the P-only branch. The instability of the P-only branch occurs continuously
at $a_m^{\rm eff}=0$, where $m_0=0$; minimizing Eq.~\ref{eq:SI_feff_m} for
$a_m^{\rm eff}<0$ gives
\begin{equation}
  m_0^2 = -\frac{a_m^{\rm eff}}{b_m^{\rm eff}},
  \label{eq:SI_m0_via_eff}
\end{equation}
identical to Eq.~\ref{eq:SI_m0_full}.

Evaluating $f_{\rm red}(P)$ on the P-only branch gives
$f_{P\text{-only}}=-a_P^2/(4b_P)$, the constant term of $f_{\rm eff}(m)$.
Minimizing Eq.~\ref{eq:SI_feff_m} gives the free energy of the fully coupled
branch relative to this value,
\begin{equation}
  f_{\rm full}-f_{P\text{-only}}
  = -\frac{\left(a_m^{\rm eff}\right)^2}{4\,b_m^{\rm eff}},
  \label{eq:SI_fdiff}
\end{equation}
which vanishes at $a_m^{\rm eff}=0$ and is negative for $a_m^{\rm eff}<0$
whenever $D>0$: beyond the instability of the P-only branch, the fully
coupled branch is thermodynamically favored.

\subsection{Hessian and stability}
\label{sec:SI_stability_Hessian}

The Hessian of the free-energy density at a general point $(m,P)$ is
\begin{equation}
  H =
  \begin{pmatrix}
    a_m+3b_m m^2-gP^2 & -2gmP\\
    -2gmP & a_P+3b_P P^2-gm^2
  \end{pmatrix}.
  \label{eq:SI_Hessian}
\end{equation}
At the P-only stationary point $m=0$, $P^2=-a_P/b_P$, the off-diagonal element
vanishes and the eigenvalues are $a_m^{\rm eff}$ and $-2a_P=2b_PP^2$. The
P-only stationary point is a local minimum when $a_m^{\rm eff}>0$ and
$-2a_P>0$; since this branch has $a_P<0$ and $b_P>0$, the second condition is
already satisfied. Global boundedness of the quartic free energy additionally
requires $b_m>0$, $b_P>0$, and $D>0$ (Eq.~\ref{eq:SI_D_positive}).

At the fully coupled stationary point $(m_0,P_0)$, using the stationarity
conditions to eliminate $a_m$ and $a_P$ from the diagonal entries of
Eq.~\ref{eq:SI_Hessian} gives
\begin{equation}
  H_{\rm full}
  =
  2\begin{pmatrix}
    b_m m_0^2 & -g m_0 P_0\\[2pt]
    -g m_0 P_0 & b_P P_0^2
  \end{pmatrix},
  \label{eq:SI_Hessian_full}
\end{equation}
with
\begin{equation}
  \det H_{\rm full}
  = 4m_0^2P_0^2\,D,
  \qquad
  \operatorname{Tr}H_{\rm full} = 2\left(b_m m_0^2+b_P P_0^2\right) > 0
  \label{eq:SI_detH_full}
\end{equation}
whenever $b_m,b_P>0$ and $m_0^2,P_0^2>0$.

As $a_m^{\rm eff}$ decreases through zero, the P-only minimum loses one
positive curvature and becomes unstable for $a_m^{\rm eff}<0$; the fully
coupled branch is then the stable local minimum, provided its two squared
amplitudes are positive and $D>0$.

\newpage
\section{Parameter Identifiability and Rayleigh Broadening}
\label{sec:numerics}

\subsection{Identifiable parameter combinations}
\label{sec:calibration}

The structural threshold observed in the Rayleigh/IR data~\cite{Sandeep2024},
$a_m^{\rm eff}=0$, written using
Eq.~\ref{eq:SI_aeff_closure} and Eq.~\ref{eq:SI_aP}, solves for
\begin{equation}
  \Omega_s^2(\Delta,T) = A_T\,T + A_\Delta\,\Delta^2 + B_s,
  \label{eq:SI_Omegas_ATAdBs}
\end{equation}
with
\begin{equation}
  A_T \equiv \frac{\alpha_T}{\alpha_\Omega},
  \qquad
  A_\Delta \equiv \frac{\alpha_\Delta}{\alpha_\Omega},
  \qquad
  B_s \equiv \frac{b_P\,a_m/g - \alpha_T T_0}{\alpha_\Omega}.
  \label{eq:SI_ATAdBs_def}
\end{equation}
For comparison, the primary boundary (Eq.~\ref{eq:SI_OmegaC}) is
\begin{equation}
  \Omega_c^2(\Delta,T) = A_T\,T + A_\Delta\,\Delta^2 - A_T\,T_0,
  \label{eq:SI_Omegac_ATAdBs}
\end{equation}
so that
\begin{equation}
  \Omega_s^2 - \Omega_c^2 = \frac{b_P\,a_m}{g\,\alpha_\Omega}.
  \label{eq:SI_Omega_offset}
\end{equation}

$A_T$ is fixed by the temperature slope of the structural threshold,
$A_\Delta$ by its detuning curvature, and $B_s$ by its intercept. $T_0$ and
$b_Pa_m/(g\alpha_\Omega)$ cannot be separated from Rayleigh-threshold data
alone, since they enter only through $B_s$; separating them requires an
independent observable that constrains the primary $a_P=0$ locus. The overall
GL mass and field amplitudes retain a normalization freedom: once
$\alpha_\Omega$, $a_m$, and $b_P$ are fixed by convention and $T_0$ is
selected, $g$ is assigned to reproduce $B_s$. $\kappa$ requires direct
correlation-length data for an absolute calibration, and $b_m$ does not affect
the threshold locus.

For the parameters of Table~1 of the main text, $B_s\approx1.03\times10^3\,
\mathrm{cm}^{-2}$, giving the structural threshold near
$\Omega^2=10^3\,\mathrm{cm}^{-2}$ at $T=0$ and $\Delta=0$. $A_T$ controls the
temperature slope, $A_\Delta$ the detuning curvature, and $B_s$ the intercept;
neither $b_m$ nor $\kappa$ changes the uniform threshold locus.

Table~\ref{tab:SI_calibration} summarizes this identifiability structure.

\begin{table}[htbp]
  \caption{Identifiability of the structural/Rayleigh threshold combinations.}
  \label{tab:SI_calibration}
  \footnotesize
  \begin{ruledtabular}
  \begin{tabular}{@{}lp{4.3in}@{}}
    Quantity & Identifiability \\
    \hline
    $A_T=\alpha_T/\alpha_\Omega$ & fixed by the temperature slope of the
      structural threshold \\
    $A_\Delta=\alpha_\Delta/\alpha_\Omega$ & fixed by the detuning curvature
      of the structural threshold \\
    $B_s$ (Eq.~\ref{eq:SI_ATAdBs_def}) & fixed by the threshold intercept \\
    $T_0$, $b_Pa_m/(g\alpha_\Omega)$, the absolute field normalization,
      $b_m$, $\kappa$ & not separately determined by the structural threshold
      alone \\
  \end{tabular}
  \end{ruledtabular}
\end{table}

\begin{figure}[htbp]
  \centering
  \includegraphics[width=0.56\textwidth]{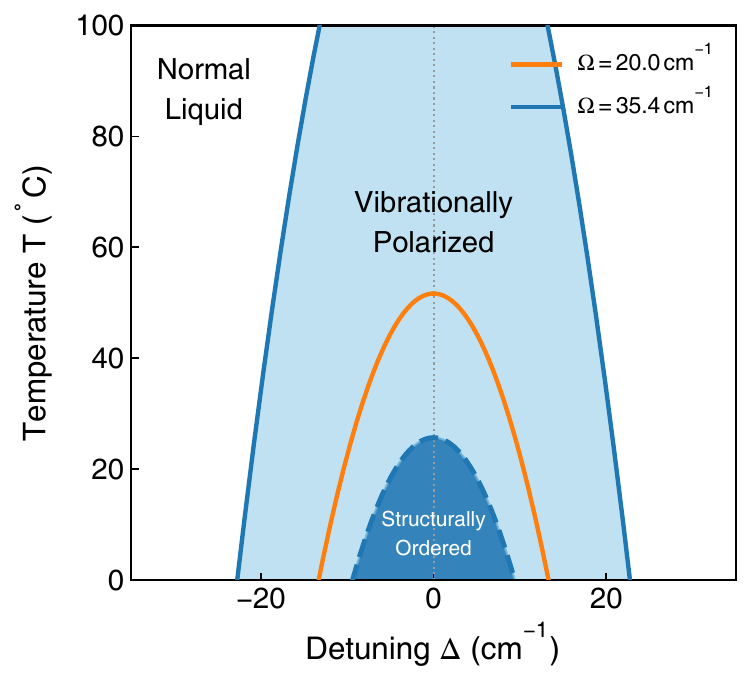}
  \caption{%
    \textbf{Detuning--temperature projection of the GL phase diagram.}
    Solid curves: the model primary boundary $a_P=0$ for the selected $T_0$;
    its absolute location is not separately determined unless $T_0$ is
    independently constrained. Dashed curves: the structural threshold
    $a_m^{\rm eff}=0$, determined by $A_T$, $A_\Delta$, and $B_s$
    (Eq.~\ref{eq:SI_Omegas_ATAdBs}); the separation between the solid and
    dashed curves is $b_Pa_m/(g\alpha_\Omega)$ (Eq.~\ref{eq:SI_Omega_offset}).
    Shaded regions distinguish the two sides of the dashed curve for
    $\Omega=\sqrt{1250}\simeq35.4\,\mathrm{cm}^{-1}$; for
    $\Omega=20\,\mathrm{cm}^{-1}$ the dashed curve lies below
    $0\,^\circ$C over the plotted range.
  }
  \label{fig:SI_delta_T_phase}
\end{figure}

\subsection{Broadening of the Rayleigh crossover}
\label{sec:SI_readout_crossing}

In the ideal limit of a spatially uniform sample and infinite resolution, the
Rayleigh visibility is a step,
\begin{equation}
  S_R^{(0)}(a_m^{\rm eff})=\Theta(-a_m^{\rm eff}).
  \label{eq:SI_step_readout}
\end{equation}
Local variations in cavity thickness, concentration, detuning, temperature,
and structural environment shift the local mass, $a_m^{\rm eff}\to
a_m^{\rm eff}+\eta$, and the measured visibility is the disorder-averaged
step,
\begin{equation}
  S_R(a_m^{\rm eff})
  =
  \int d\eta\,p(\eta)\,\Theta[-(a_m^{\rm eff}+\eta)].
  \label{eq:SI_readout_average}
\end{equation}
A logistic $p(\eta)$ of width $\sigma_a$ gives Eq.~6 of the main text.
$\sigma_a$ broadens the crossover; the midpoint remains at $a_m^{\rm eff}=0$;
$\sigma_a$ does not enter the GL free energy.

\begin{figure}[htbp]
  \centering
  \includegraphics[width=0.56\textwidth]{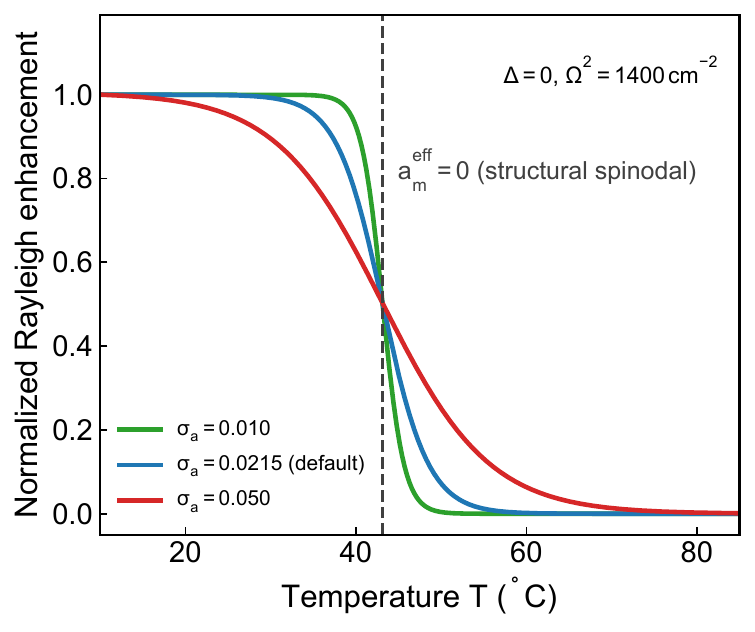}
  \caption{%
    \textbf{Effect of the broadening width $\sigma_a$ on the normalized
    Rayleigh crossover.}
    Normalized Rayleigh enhancement versus temperature at $\Delta=0$,
    $\Omega^2=1400\,\mathrm{cm}^{-2}$, for three values of $\sigma_a$; each
    curve is normalized by its maximum. Changing $\sigma_a$ changes the width
    of the crossover but not its midpoint, $a_m^{\rm eff}=0$ (dashed vertical
    line).
  }
  \label{fig:SI_sigma_a}
\end{figure}

\newpage
\section{Static and Dynamic Structural Correlations}
\label{sec:SI_structure_correlations}

$S(q)$ denotes the coarse-grained, long-wavelength structure factor
associated with the structural field $m(\mathbf r)$.

On the stable side of the structural instability ($a_m^{\rm eff}>0$), the
quadratic free energy for $m$ is, in Fourier space,
\begin{equation}
  F_m^{(2)}
  =
  \frac{1}{2}
  \int\frac{d^3q}{(2\pi)^3}\,
  A_q\,|m_q|^2,
  \qquad
  A_q\equiv a_m^{\rm eff}+\kappa q^2,
  \label{eq:SI_Fm_dynamic}
\end{equation}
with overdamped, Model-A-like dynamics~\cite{HohenbergHalperin1977},
\begin{equation}
  \partial_t m_q(t)
  =
  -\Gamma_m A_q\,m_q(t)+\eta_q(t),
  \qquad
  \langle \eta_q(t)\eta_{-q}(t')\rangle
  =
  2\Gamma_m k_\mathrm{B}T\,\delta(t-t').
  \label{eq:SI_TDGL_nonconserved}
\end{equation}
Solving Eq.~\ref{eq:SI_TDGL_nonconserved} gives the time correlation function
\begin{equation}
  C(q,t)\equiv\langle m_q(t)m_{-q}(0)\rangle
  =
  \frac{k_\mathrm{B}T}{A_q}\,e^{-|t|/\tau_q},
  \qquad
  \tau_q^{-1}=\Gamma_m A_q,
  \label{eq:SI_Cqt}
\end{equation}
and its Fourier transform is the dynamic structure factor,
\begin{equation}
  S(q,\omega)\equiv\int_{-\infty}^{\infty}dt\,e^{i\omega t}C(q,t)
  =
  \frac{2\Gamma_m k_\mathrm{B}T}{\omega^2+\Gamma_m^2 A_q^2}.
  \label{eq:SI_Sqomega}
\end{equation}
The static structure factor is the frequency integral,
\begin{equation}
  S(q) \equiv C(q,0) = \int_{-\infty}^{\infty}\frac{d\omega}{2\pi}\,S(q,\omega)
  =
  \frac{k_\mathrm{B}T}{A_q}
  =
  \frac{S(0)}{1+q^2\zeta^2},
  \qquad
  S(0)=\frac{k_\mathrm{B}T}{a_m^{\rm eff}},
  \qquad
  \zeta=\sqrt{\frac{\kappa}{a_m^{\rm eff}}},
  \label{eq:SI_Sq}
\end{equation}
valid only for $a_m^{\rm eff}>0$. With $S_{\rm out}(0)=k_\mathrm{B}T/a_m$ and
$\zeta_{\rm out}=\sqrt{\kappa/a_m}$,
\begin{equation}
  \frac{S(0)}{S_{\rm out}(0)}
  =
  \frac{a_m}{a_m^{\rm eff}}
  =
  \left(\frac{\zeta}{\zeta_{\rm out}}\right)^2.
  \label{eq:SI_S0_ratio}
\end{equation}

Fourier transforming Eq.~\ref{eq:SI_Sq} gives the real-space field
correlation,
\begin{equation}
  C(r)\equiv\langle\delta m(\mathbf r)\delta m(\mathbf 0)\rangle
  =
  \frac{k_\mathrm{B}T}{4\pi\kappa}\,\frac{e^{-r/\zeta}}{r},
  \label{eq:SI_Cr}
\end{equation}
which maps onto the pair-correlation function
\begin{equation}
  h(r)\equiv g(r)-1 \sim A\,\frac{e^{-r/\zeta}}{r}
  \label{eq:SI_h_tail}
\end{equation}
up to a nonuniversal contrast amplitude $A$; the model does not reconstruct
the molecular-scale $g(r)$, only the long-distance decay form and $\zeta$.

At $q=0$,
\begin{equation}
  S(0)=\frac{k_\mathrm{B}T}{a_m^{\rm eff}},
  \qquad
  \tau_0=\frac{1}{\Gamma_m a_m^{\rm eff}},
  \qquad
  S(0,0)=\frac{2k_\mathrm{B}T}{\Gamma_m(a_m^{\rm eff})^2}.
  \label{eq:SI_critical_scaling}
\end{equation}
As $a_m^{\rm eff}\to0^+$, the integrated intensity and relaxation time grow as
$1/a_m^{\rm eff}$, and the zero-frequency spectral density grows as
$1/(a_m^{\rm eff})^2$.

If $m$ is instead a strictly conserved density field, the dynamics is
diffusive, with relaxation rate
\begin{equation}
  \tau_q^{-1}=M q^2 A_q;
  \label{eq:SI_conserved_rate}
\end{equation}
the static structure factor $S(q)=k_\mathrm{B}T/A_q$ is unchanged, but the
extra factor of $q^2$ modifies the hydrodynamic small-$q$ linewidth.

\paragraph{Parameters used in Fig.~4.}
All curves have $a_m^{\rm eff}>0$ and $\kappa=1$; $q$ is in units of
$\zeta_{\rm out}^{-1}$ and $r$ in units of $\zeta_{\rm out}$, with
$r_{\rm min}=0.3\,\zeta_{\rm out}$. All real-space curves share the same
nonuniversal amplitude $A$, which cancels in $h(r)/h_{\rm out}(r_{\rm min})$.
The value $a_m^{\rm eff}=\sigma_a$ used for the near-spinodal curve is only a
plotting choice of a small positive reference mass; it does not imply a
thermodynamic relation between the OZ mass and the broadening width
$\sigma_a$.


\bibliography{ref}